\documentclass[%
reprint,
superscriptaddress,
amsmath,amssymb,
aps,
prb,
]{revtex4-2}

\usepackage{graphicx}% Include figure files
\usepackage{dcolumn}% Align table columns on decimal point
\usepackage{bm}% bold math
\usepackage{hyperref}% add hypertext capabilities
\hypersetup{
    colorlinks = true,
    linkcolor  = {[rgb]{0.0, 0.0, 0.6}},   % dark blue
    citecolor  = {[rgb]{0.0, 0.0, 0.6}},
    urlcolor   = {[rgb]{0.0, 0.0, 0.6}}
}
\usepackage{array} % add this to your preamble if not already present
\usepackage[mathlines]{lineno}% Enable numbering of text and display math
\usepackage{newtxtext,newtxmath}
\usepackage{tabularx} % Include this in your preamble
\usepackage{booktabs} % Optional: for professional table rules

\begin{document}

\preprint{APS/123-QED}

\title{
Stability and Nonlinear Dynamics
of Field-Driven Extended Domain Walls with Interfacial Dzyaloshinskii–Moriya Interaction
}% Force line breaks with \\
% \author{Ann Author}
%  % \altaffiliation[Also at ]{Physics Department, XYZ University.}%Lines break automatically or can be forced with \\
%  % \email{Second.Author@institution.edu}
% \affiliation{%
%  Authors' institution and/or address\\
%  This line break forced with \textbackslash\textbackslash
% }%
\author{E. F. Pusiol}
\affiliation{Centro At\'omico Bariloche, CNEA-CONICET, 
(R8402AGP) San Carlos de Bariloche, Río Negro, Argentina}

\author{Alberto Rosso}%
\affiliation{%
Université Paris-Saclay, CNRS, LPTMS, 91405, Orsay, France.
}%

\author{A. B. Kolton}
\affiliation{Centro At\'omico Bariloche, CNEA-CONICET, 
(R8402AGP) San Carlos de Bariloche, Río Negro, Argentina}

\affiliation{Instituto Balseiro, UNCUYO,  
(R8402AGP) San Carlos de Bariloche, Río Negro, Argentina
}

\date{\today}% It is always \today, today,

\begin{abstract}

We analyze the stability and non-linear dynamics of extended domain walls (DWs) with Dzyaloshinskii-Moriya interaction (DMI) driven by an external magnetic field in the clean limit (i.e., in the absence of disorder and thermal fluctuations). 
We show that for moderate DMI strengths, the dynamics in both the low- and high-field regimes are accurately described by a rigid-wall Walker model that neglects flexural degrees of freedom. 
At intermediate fields, however, the system undergoes a transition into a non-trivial spatio-temporally chaotic regime characterized by dynamically roughened domain walls displaying single- or multi-valued profiles. 
The phase boundaries delimiting this regime in the magnetic field--DMI phase space are computed using a linear stability analysis based on Floquet theory applied to the extended domain wall, explicitly accounting for DMI and internal phase dynamics. 
Our theoretical analysis and predicted phase diagram are fully supported by numerical micromagnetic simulations solving the full Landau-Lifshitz-Gilbert equation.

\end{abstract}

\maketitle

\section{\label{sec:Intro}Introduction}

Magnetic domain walls (DWs) are fundamental topological excitations in ordered media and constitute the active elements of a wide range of spintronic devices~\cite{Hubert1998,Beach2008}. Their controlled manipulation by magnetic fields and electric currents has motivated decades of theoretical and experimental research, driven by both the fundamental physics of nonequilibrium magnetization dynamics and the development of memory and logic devices based on domain-wall motion~\cite{Parkin2008,Allwood2005}. In ultrathin magnetic heterostructures, interfacial Dzyaloshinskii–Moriya interaction (DMI) has emerged as a key ingredient, stabilizing chiral Néel domain walls, magnetic skyrmions, and other topological spin textures~\cite{Camley2023,Fertetal2013,Mallick2024,kishan2025_skyrmion_review}. Besides determining their equilibrium structure, DMI strongly influences the response of these textures to magnetic fields and electric currents~\cite{thiaville2012,krizakova2019}, making it a central ingredient in modern chiral spintronics.

Despite these advances, most theoretical descriptions of driven DWs rely on rigid one-dimensional models in which the wall remains flat and its dynamics is described by a small set of collective coordinates~\cite{thiaville2012}. This approximation successfully captures the Walker breakdown~\cite{Schryer1974} and many aspects of chiral DW motion, but necessarily neglects spatial degrees of freedom along the wall. Under sufficiently strong driving, extended DWs may lose their morphological stability, leading to wall corrugation and the subsequent nucleation and annihilation of vertical Bloch lines (VBLs), giving rise to complex spatiotemporal dynamics~\cite{Slonczewski1972,thevenard2011domain,gourdon2013domain,Yoshimura2016,Pusiol2026}. How these extended instabilities evolve in the presence of DMI remains largely unexplored.

In this work, we investigate how DMI influences the stability and nonlinear dynamics of field-driven extended DWs. Building on our previous study of the DMI-free case~\cite{Pusiol2026}, we determine how DMI modifies the stability of flat walls and reorganizes the resulting dynamical regimes. Combining linear stability analysis, Floquet theory, and numerical simulations, we identify and characterize phase-locked and precessional regimes. Our results are summarized in a dynamical phase diagram as a function of the out-of-plane magnetic field and DMI strength, extending the phase diagram previously obtained in the absence of DMI~\cite{Pusiol2026}.

The remainder of the paper is organized as follows. In Sec.~\ref{sec:mumaxphenomenology}, we present the main phenomenology of field-driven extended DWs in the presence of interfacial DMI based on full micromagnetic simulations. Motivated by these observations, we introduce in Sec.~\ref{sec:reducedmodel} a DW model that captures the coupled translational and internal dynamics of extended walls; furthermore, the solutions for rigid DWs are described as a function of the DMI strength. In Sec.~\ref{sec:linearstability}, based on those solutions, we perform a linear stability analysis of their flexural modes, allowing us to determine the characteristic fields that define the dynamical phase diagram. In Sec.~\ref{sec:phasediagram}, we construct and discuss the resulting dynamical phase diagram. Finally, Sec.~\ref{sec:conclusions} summarizes our main conclusions.

\section{Micromagnetic simulations: phenomenology of driven extended DWs}
\label{sec:mumaxphenomenology}

As a base model, we assume that the magnetization dynamics are governed by the Landau-Lifshitz-Gilbert (LLG) equation
\begin{equation}
    \frac{\partial \mathbf{m}}{\partial t} = -\gamma\,\mathbf{m}\times\mathbf{H}_{\text{eff}}
    + \alpha\,\mathbf{m}\times\frac{\partial \mathbf{m}}{\partial t},
    \label{eq:LLG}
\end{equation}
where $\gamma=\gamma_{0}\mu_{0}$, $\gamma_{0}$ is the gyromagnetic ratio, $\mu_{0}$ the vacuum permeability, $\alpha$ is the Gilbert damping constant, and
$\mathbf{H}_{\text{eff}} = -(1/\mu_0 M_s)\,\delta E/\delta\mathbf{m}$
is the effective field derived from the total energy $E = \int \mathcal{E}\,d\mathbf{r}$, with energy density
\begin{align}
    \mathcal{E} =\;
    &-\mu_0 M_s \mathbf{m}\cdot\mathbf{H}
    + A_{\text{ex}}(\nabla\mathbf{m})^2
    - \frac{\mu_{0}M_s}{2}\mathbf{m}\cdot\mathbf{H}_{d}
    - K_u\, m_z^2 \nonumber \\
    &+ D\left[m_z\,(\nabla\cdot\mathbf{m})-(\mathbf{m}\cdot\nabla) m_z\right].
    \label{eq:energy_density}
\end{align}
The terms in Eq.~\eqref{eq:energy_density} represent, respectively, the Zeeman coupling to the externally applied field $\mathbf{H}$ (with $\mathbf{B} = \mu_0\mathbf{H}$), the Heisenberg exchange, the magnetostatic interaction with demagnetizing field $\mathbf{H}_d$, the perpendicular magnetic anisotropy (PMA) with constant $K_u$, and the interfacial DMI with strength $D$. Here $M_s$ is the saturation magnetization, and $A_{\text{ex}}$ is the exchange stiffness constant.

\begin{figure}[]
    \centering
    \includegraphics[]{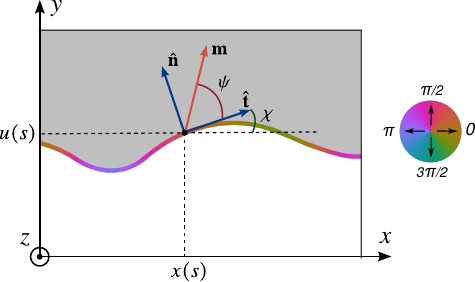}
    \caption{Schematic of the extended DW interface. The white (gray) region denotes the $+\hat{z}$ ($-\hat{z}$) domain. The interface is colored by the in-plane magnetization angle $\phi$ measured from the $x$ axis (colorwheel at right). At each point, $\hat{\mathbf{t}}$ and $\hat{\mathbf{n}}$ denote the local tangent and normal, $\chi$ is the tangent's angle from horizontal, and $\psi$ is the angle between the magnetization $\mathbf{m}$ and $\hat{\mathbf{t}}$.
    }
    \label{fig:uphi}
\end{figure}

To illustrate the nontrivial dynamics of DWs with an internal precessional degree of freedom in the presence of DMI, we numerically integrate Eq.~\eqref{eq:LLG} using \textsc{mumax\textsuperscript{3}}~\cite{vansteenkiste2014design}, starting from a slightly perturbed flat DW, and monitor its convergence to a steady state of uniform propagation (see Fig.~\ref{fig:uphi}). The DW interface is, on average, oriented along $\hat{x}$, with propagation along $\hat{y}$ for a positive applied field. Simulations are performed at zero temperature and in the absence of disorder. To be concrete we consider a $\text{Pt/Co/GdO}_x$ multilayer with the following material parameters~\cite{Pham2016, krizakova2019}: $\alpha = 0.15$, $A_{\text{ex}} = 1.6\times10^{-11}$ J/m, $M_{s} = 1.26\times10^{6}$ A/m, effective anisotropy $K_{0}=K_{u}-\mu_{0}M^{2}_{s}/2 = 0.41\times10^{6}$ J/m$^3$, and $D = 0.841$ mJ/m$^2$. The simulated film has dimensions $1024\times1024\times1$ nm$^3$, discretized on a uniform grid with cell size $2\times2\times1$ nm$^3$, and periodic boundary conditions are imposed along the $x$ direction. 

\begin{figure*}[]
   \centering
   \includegraphics[]{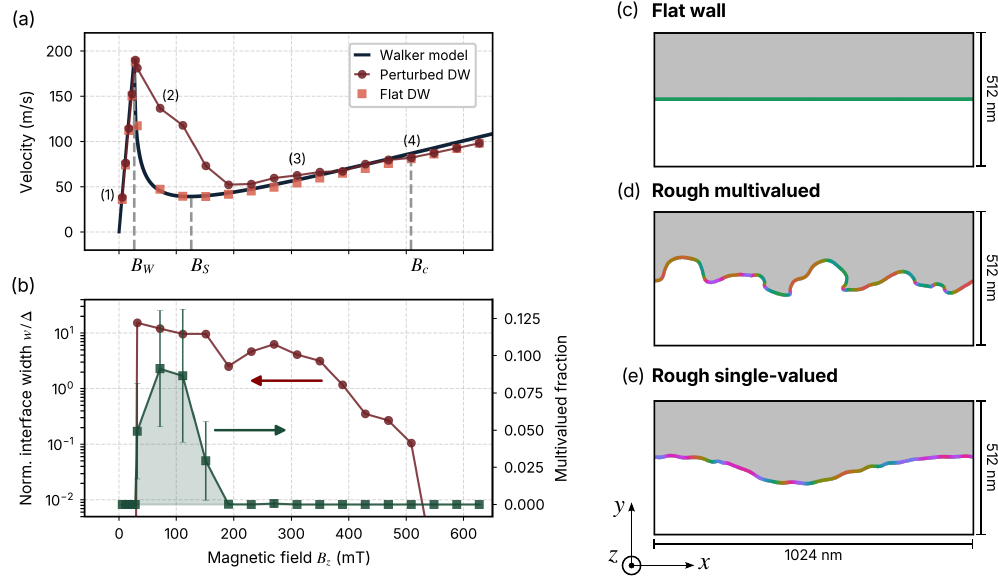}
   \caption{Field-driven dynamics of an extended DW in the presence of DMI.
   (a) DW velocity as a function of the driving field $B_z$ of a perturbed extended DW (red dots), compared with the prediction of the Walker model [solid line, Eq.~\eqref{eq:u_dot_walker}] and with micromagnetic simulations of an unperturbed flat DW (squares).
   (b) Normalized interface width $w/\Delta$ [Eq.~\eqref{eq:width}, dots left axis] and multivalued fraction $f_m$ [Eq.~\eqref{eq:fm}, squares right axis] as a function of $B_z$, quantifying the roughness and the degree of overhang proliferation of the DW, respectively.
   (c)--(e) Steady-state magnetic configurations at long times, for the points marked (1), (2), and (3) in panel (a), respectively, corresponding to the non-precessional regime (c) and to the rough phase (d), (e). The interface color encodes the local internal magnetization angle $\phi(x,t)$ according to the circular map shown in Fig.~\ref{fig:uphi}.
}
   \label{Fig:f_sec1}
\end{figure*}

Under a uniform out-of-plane magnetic field $\mathbf{B} = B_{z}\hat{z}$, we collected magnetic configurations at a sufficiently high temporal sampling rate to resolve the internal magnetization dynamics of the DW. For each magnetic configuration, we extract the DW interface as a parametrized curve $\mathbf{r}(s,t) = (x(s,t), u(s,t))$, defined implicitly by $\mathbf{m}(\mathbf{r},t)\cdot\hat{z}=0$, where the parameter $s$ indexes points along the interface (see Fig.~\ref{fig:uphi}). This parametrization naturally accommodates overhangs, since $x(s,t)$ need not be single-valued: DMI is known to favor their formation~\cite{Pham2016, krizakova2019, Yoshimura2016}, and in such cases a given $x$ is crossed by the interface multiple times as $s$ increases. We denote by 
\begin{equation}
    \langle f \rangle_{s}
    (t) 
    \equiv P(t)^{-1}\int_{s=0}^{P(t)} f(s,t)ds
\end{equation}
the spatial average of a quantity $f(s,t)$ over the interface at fixed $t$ with perimeter $P(t)$. The DW position is then
\begin{equation}
    u(t) \equiv \langle u(s,t) \rangle_{s},
\end{equation}
and the interface width is
\begin{equation}
    w(t) = \sqrt{\langle u^{2}\rangle_{s} - \langle u \rangle_{s}^{2}}.
    \label{eq:width}
\end{equation}
We further define the multivalued fraction $f_{m}$ (relative to the constant $y$ base plane)
as the fraction of interface points at which the local tangent $\hat{\mathbf{t}}(s,t)$ points against the $x$ direction,
\begin{equation}
    f_{m}(t) = \big\langle\, \Theta\;\!\big(-\hat{\mathbf{t}}(s,t)\cdot\hat{x}\big)\big\rangle_{s},
    \label{eq:fm}
\end{equation}
where $\Theta$ is the Heaviside step function, such that $f_{m}=0$ for a flat or single-valued interface (in which case we can parametrize $ds\equiv dx$).

In Figs.~\ref{Fig:f_sec1}(a) and~\ref{Fig:f_sec1}(b) we show how typical transport and geometrical properties of a driven extended DW evolve as a function of $B_{z}$. Panel~(a) shows the DW velocity $\langle \dot u \rangle$, panel~(b) shows the interface width $\langle w \rangle$ normalized by the DW width $\Delta=\sqrt{A_{\text{ex}}/K_0}$ (left axis) and the overhang fraction $\langle f_m \rangle$ (right axis), where $\langle\cdot\rangle$ denotes a time average over the steady state. Also, for comparison, in Fig.~\ref{Fig:f_sec1}(a) we show results from micromagnetic LLG simulations of an unperturbed, flat chiral DW, and the prediction of the Walker model~\cite{thiaville2012},
\begin{equation}
    \langle \dot{u} \rangle =
    \begin{cases}
        \dfrac{\Delta}{\alpha}\gamma_{0}\,B_{z}, \quad B\leq B_{W}, \\
        \\
        \dfrac{\Delta}{\alpha} \left[ \gamma_{0}\,B_{z}- \dfrac{2\pi}{T(B_{z})}\right], \quad B>B_{W},
    \end{cases}
    \label{eq:u_dot_walker}
\end{equation}
which treats the DW as a particle-like object with no internal spatial degrees of freedom, with $T$ the field-dependent precession period, Eq.~\eqref{eq:period_general}.

In the field range $B_{z} = 0\text{--}28.25$~mT, for the chosen micromagnetic parameters, the interface moves rigidly in the \emph{non-precessional} regime (also called \emph{stationary regime}): the internal magnetization angle $\phi$ is pinned at a field-dependent value, and the velocity increases linearly with $B_{z}$. As an example, Fig.~\ref{Fig:f_sec1}(c) shows the equilibrium configuration obtained at the point $(B_z,v)$ marked as (1) in Fig.~\ref{Fig:f_sec1}(a), with the interface color encoding the equilibrium angle $\phi$. Its dependence on the driving field and the DMI strength is discussed in Sec.~\ref{sec:fixed_points}. In terms of the DW shape, the interface remains essentially flat, with $w/\Delta \ll 1$, and the velocity closely matches the prediction of the Walker model and the micromagnetic results for the flat interface.

This regime terminates at the \textit{Walker field} $B_{W}$, above which $\phi$ begins to precess continuously, setting the onset of the \emph{precessional regime}. In the Walker model, this regime exhibits a region of negative differential mobility immediately above $B_{W}$, with the velocity reaching a local minimum at the Slonczewski field $B_{S}$. Heuristic arguments have long suggested that this region is associated with morphological instabilities of extended DWs~\cite{Slonczewski1972, krizakova2019, Garcia2021}.

\begin{figure}[]
    \centering
    \includegraphics[]{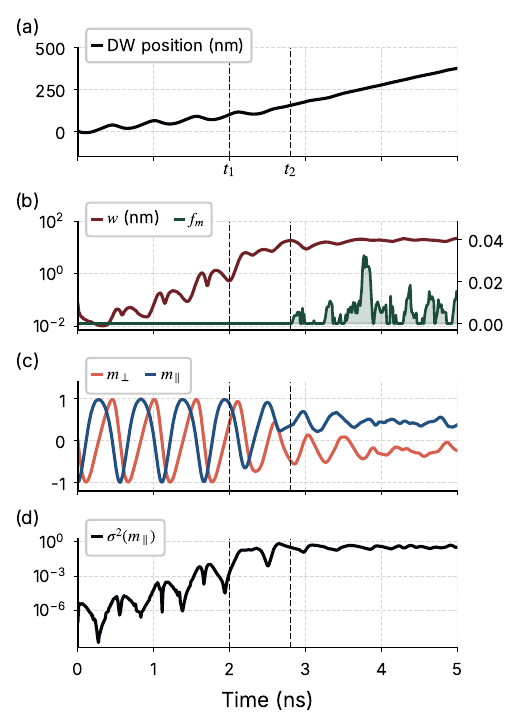}
    \caption{Short-time transient dynamics of the DW in the corrugated phase, at the driving field $B_z = 71.4\,\mathrm{mT}$ [point (2) in Fig.~\ref{Fig:f_sec1}(a)].
    (a) DW position $u(t)$.
    (b) Interface width $w(t)$ (left axis) and multivalued fraction $f_{m}(t)$ (right axis).
    (c) Parallel and perpendicular components of the magnetization measured from the local tangent.
    (d) Spatial variance of the parallel magnetization component, $\sigma_{\parallel}^{2}(t)$.}
    \label{fig:transient}
\end{figure}

Beyond this interpretation, we find that extended DWs remain unstable even for fields above $B_{S}$: \emph{rough} configurations emerge from $B_{W}$ up to a critical field $B_{c} = 508.5$~mT. This dynamical regime presents significant deviations from rigid, flat-wall motion, exhibiting high velocity, large geometrical distortions, and a spatially non-uniform internal angle. Figures~\ref{Fig:f_sec1}(d) and~\ref{Fig:f_sec1}(e) illustrate magnetic configurations, extracted at long times, at the points marked (2) and (3) in Fig.~\ref{Fig:f_sec1}(a), respectively. Near the Walker field, the proliferation of overhangs is evidenced by the sharp increase in the multivalued fraction [Fig.~\ref{Fig:f_sec1}(b)]. As the driving field is increased further, the interface progressively converges to a single-valued fluctuating random function. 

Finally, at high fields the precessional dynamics re-stabilizes the flat wall configuration: at the point marked (4) in Fig.~\ref{Fig:f_sec1}(a), for example, the interface width drops significantly, towards values $w/\Delta \ll 1$, and the velocity matches that of the unperturbed flat DW. The minor discrepancies in the velocity-field curve between the simulated flat DW and the Walker model can be explained by the simplifying assumptions of the model: in particular, it treats the DW width as a material constant independent of film geometry. A proper analytical treatment of the magnetostatic energy yields a correction that depends on the thin-film thickness~\cite{skaugen2019}.

Taken together, these results indicate that roughening emerges in a finite window of driving fields, bounded below by $B_{W}$ and above by a critical field $B_{c}$, outside of which the extended DW recovers a rigid, flat-wall motion. This raises the question of how the instability responsible for the roughening process actually develops: starting from a weakly perturbed flat DW, how does the flexural distortion of the interface emerge, and how does it couple to the internal precessional dynamics of $\phi$?

To address this, we now turn to the short-time, transient dynamics of the DW at a representative field within the rough phase, corresponding to point (2) in Fig.~\ref{Fig:f_sec1}(a). Figure~\ref{fig:transient} shows, as a function of time, (a) the DW position $u(t)$, (b) the interface width $w(t)$ [Eq.~\eqref{eq:width}] on left axis and the multivalued fraction $f_{m}$ [Eq.~\eqref{eq:fm}] on right axis, and (c) the parallel and perpendicular components of the magnetization with respect to the local tangent and normal directions (see Fig.~\ref{fig:uphi}),
\begin{align}
    m_{\parallel}(t) &= \langle \mathbf{m}(s,t) \cdot \hat{\mathbf{t}}(s,t) \rangle_{s}, \\
    m_{\perp}(t) &= \langle \mathbf{m}(s,t) \cdot \hat{\mathbf{n}}(s,t) \rangle_{s}.
\end{align}
These spatially averaged projections quantify the degree of Bloch ($m_{\parallel}$) and Néel ($m_{\perp}$) character of the DW as a whole. In Fig.~\ref{fig:transient}(d) we additionally show the spatial variance of the parallel magnetization component, $\sigma^{2}_{\parallel}(t) = \langle m_{\parallel}^{2}\rangle_{s} - \langle m_{\parallel}\rangle_{s}^{2}$, quantifying the uniformity of the DW chirality.

At early times, $m_{\parallel}$ and $m_{\perp}$ oscillate coherently with constant amplitude, while both $w(t)$ and $\sigma^{2}_{\parallel}(t)$ remain small, consistent with a nearly rigid DW whose internal angle $\phi$ precesses uniformly along the interface. Beyond $t_{1}\sim2$~ns, however, the oscillation amplitude of $m_{\parallel}$ and $m_{\perp}$ begins to decay and become irregular, signaling the loss of spatial uniformity of $\phi$. The interface remains single-valued, so that $f_{m}$ stays zero, until $t_{2}\sim2.8$~ns, when a rough, multivalued interface develops. This is accompanied by an increase in the slope of $u(t)$, indicating enhanced DW mobility, together with a concurrent, roughly exponential growth of $w(t)$ and $\sigma^{2}_{\parallel}(t)$, both of which saturate on comparable timescales.

\begin{figure}[]
    \centering
    \includegraphics[]{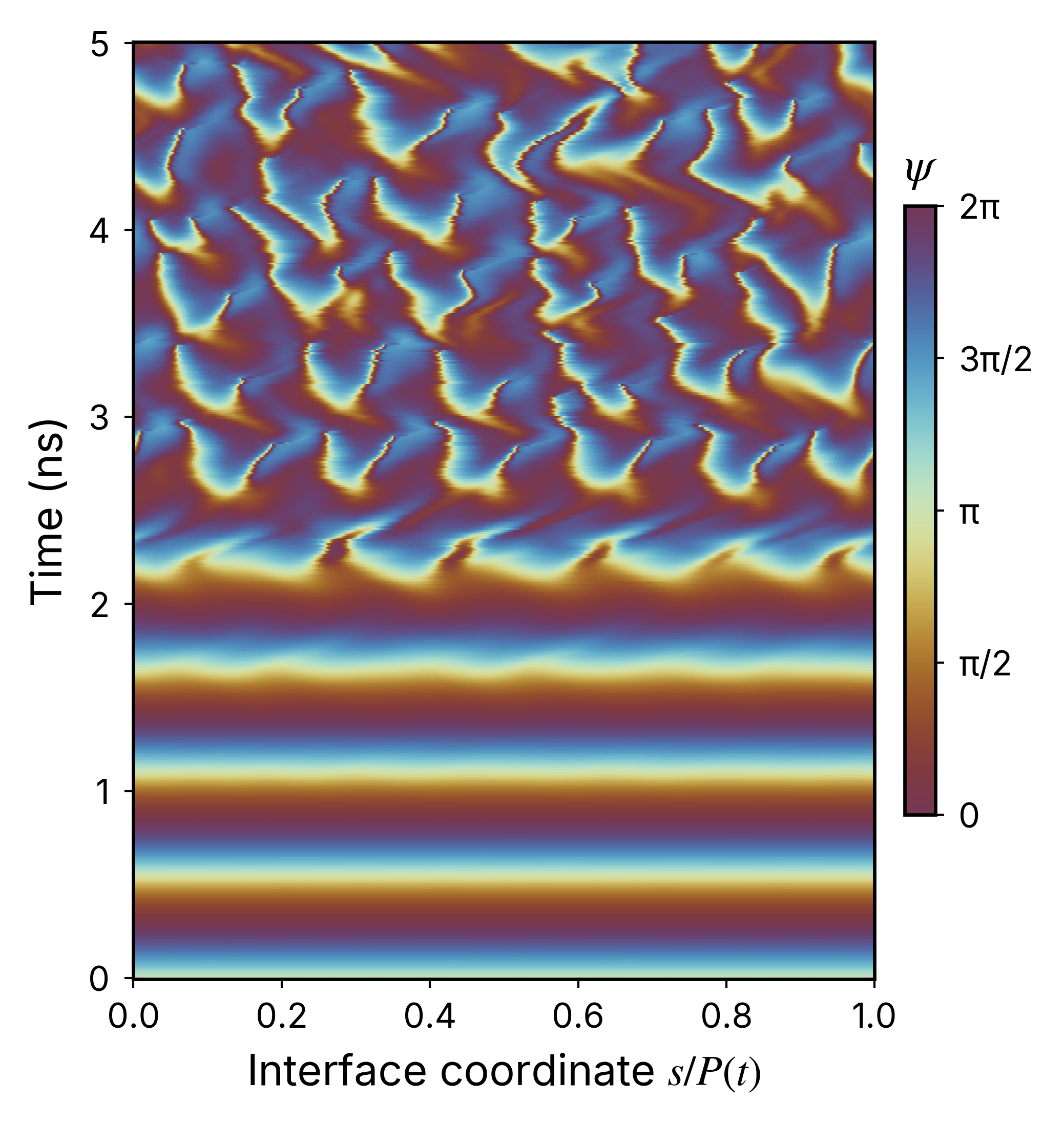}
    \caption{Space-time transient evolution of the angle $\psi$ between the local tangent and the DW magnetization, for the same driving field and time window as in Fig.~\ref{fig:transient}. 
    }
    \label{fig:space_time}
\end{figure}

In Figure~\ref{fig:space_time} we show the space-time evolution of the angle $\psi$ 
between the local tangent $\hat{\mathbf{t}}$ and the magnetization (See Fig.~\ref{fig:uphi}), for the same field and time window as in Fig.~\ref{fig:transient}. For a flat interface, $\psi=\phi$; however, in the presence of large distortions or overhangs, $\psi$ enables clear identification of topological defects (kinks in the DW magnetization angle $\phi$) along the interface. The horizontal axis indexes points along the interface, normalized by its perimeter $P(t)$. At early times, $\psi$ precesses uniformly along the interface, consistent with $w$ and $\sigma_{\parallel}^{2}$ in Figs.~\ref{fig:transient}(b)--(d). Above $t_{1}\sim2$~ns, sharp, localized discontinuities in $\psi$ emerge and propagate along the interface, consistent with the nucleation and annihilation of domain wall skyrmions~\cite{Cheng2019, krizakova2019}.

These results show that roughening emerges from a flexural instability of the flat chiral DW, in which the growth of interface width is coupled to the loss of spatial uniformity of the internal magnetization angle. This instability, driving an initially coherent DW motion to a spatiotemporal incoherent motion, is bounded in field, developing above $B_{W}$ and eventually giving way to a re-stabilized flat DW at high fields. To understand the origin of this instability, and in particular the role played by the DMI in setting its onset and extent, we develop in the next section a reduced, collective-coordinate description of the DW interface that allows us to study the linear stability of the flat chiral DW against flexural perturbations.

\section{Domain Wall model}

\label{sec:reducedmodel}
We will consider one-dimensional DWs in two-dimensional films, such that
the polar and azimuthal angles $\theta=\theta(x,y,t)$ and $\varphi=\varphi(x,y,t)$ of the unit magnetization vector $\mathbf{m} = (\sin\theta\cos\varphi,\,\sin\theta\sin\varphi,\,\cos\theta)$ are,
\begin{align}
    \theta(x,y,t) &= 2\arctan\!\left(\exp\!\left\{\frac{y - u(x,t)}{\Delta}\right\}\right),
    \label{eq:theta}\\
    \varphi(x,y,t) &= \phi(x,t),
    \label{eq:phi}
\end{align}
where $u(x,t)$ and $\phi(x,t)$ represent the local position and the internal magnetization angle of the DW, respectively. We note that this derivation treats the DW as a constant uniform-width $\Delta$ single-valued directed elastic interface described by the scalar fields $u(x,t)$ and $\phi(x,t)$ (see Fig.~\ref{fig:uphi}), thereby prohibiting overhangs. Since this description assumes a well-defined, single-valued interface, it is restricted to a linear stability analysis about the flat DW, where such an assumption holds. It therefore goes beyond standard elastic interface models described only by $u(x,t)$ by incorporating an internal extended degree of freedom $\phi(x,t)$, but its self-consistency must be checked in each regime of interest. 

In spherical coordinates, Eq.~\eqref{eq:LLG} takes the form
\begin{align}
    \sin\theta\;\dot{\varphi} &= \phantom{-}\frac{\gamma}{\mu_0 M_s}\frac{\delta E}{\delta\theta}
    + \alpha\dot{\theta}, \label{eq:LLG_theta}\\
    \dot{\theta} &= -\frac{\gamma}{\mu_0 M_s\sin\theta}\frac{\delta E}{\delta\varphi}
    - \alpha\sin\theta\;\dot{\varphi}. \label{eq:LLG_phi}
\end{align}
Substituting the ansatz~\eqref{eq:theta}--\eqref{eq:phi} in Eqs.~\eqref{eq:LLG_theta}--\eqref{eq:LLG_phi} and integrating across the DW profile (over the $y$ direction) yields coupled equations of motion for the scalar fields,
\begin{align}
    \begin{pmatrix} \dot{u} \\ \dot{\phi} \end{pmatrix}
    = \frac{\gamma}{2\mu_0 M_s(1+\alpha^2)}
    \begin{pmatrix} -\alpha\Delta & 1 \\ -1 & -\alpha/\Delta \end{pmatrix}
    \begin{pmatrix} \delta\sigma/\delta u \\ \delta\sigma/\delta\phi \end{pmatrix},
    \label{eq:u_dot_phi_dot}
\end{align}
where the $y$-integrated energy per area unit $\sigma$ functional derivatives are
\begin{align}
    \frac{\delta\sigma}{\delta u} = 
    &-2\mu_0 M_s H_z 
    - \frac{4A_{\text{ex}}}{\Delta}\,u_{xx}
    + 4\Delta K_d\phi_x\cos2\phi \nonumber \\
    &- \pi D\phi_x\sin\phi\,, \label{eq:delta_e_u}\\
    \frac{\delta\sigma}{\delta\phi} = &-4A_{\text{ex}}\Delta\,\phi_{xx}
    + 2\Delta K_d\left(\sin2\phi 
    - 2u_x\cos2\phi\right) \nonumber \\
    &+ \pi D\!\left(u_x\sin\phi + \cos\phi\right), \label{eq:delta_e_phi}
\end{align}
where the subscript denotes $\partial_{x}$. To derive these expressions, we consider the energy densities from Eq.~\eqref{eq:energy_density} considering the in-plane magnetostatic energy density
\begin{equation}
    \mathcal{E}_{d} = K_{d}(\mathbf{m}\cdot\mathbf{n})^2 =K_{d}\;\sin^2\theta\,\sin^2\psi,
    \label{eq:dens_demag}
\end{equation}
with $K_d \equiv \mu_0 M_s^2 N_n / 2$, $N_n$ the demagnetizing factor long the local DW normal $\hat{\mathbf{n}}(x)$, and $\psi=\phi-\chi$ the angle of $\mathbf{m}$ measured from the local tangent (See Fig.~\ref{fig:uphi}). The local tilt angle of the interface is defined as $\chi = \arctan(u_x)$. Assuming that the DW position can be described by a smooth displacement field $u(x,t)$, we expand to first order in $u_x$ to obtain
\begin{equation}
    \sin^2\psi\simeq\sin^2\phi - u_x\sin 2\phi.
\end{equation}
Direct inspection of Eqs.~\eqref{eq:delta_e_u}--\eqref{eq:delta_e_phi} shows $\delta\sigma/\delta u$ and $\delta \sigma/\delta \phi$ are invariant under the combined transformation $\phi\rightarrow\phi+\pi$, $D\rightarrow-D$. Consequently, if $(u(t),\phi(t))$ is a solution of Eq.~\eqref{eq:u_dot_phi_dot} for $D$, then $(u(t),\phi(t)-\pi)$ solves it for $-D$. The sign of the constant $D$ therefore selects between two branches of the same dynamics related by a $\pi$-rotation of the internal wall angle, and we restrict the analysis to $D\geq0$ without loss of generality. Furthermore, in the subsequent analysis we assume an external magnetic field applied in $+\hat{z}$ direction consistent with the DW moving in $+\hat{y}$ direction.

We nondimensionalize the model using characteristic length, time, and magnetic field scales. Distances are measured in units of $\Delta$ along the $y$-direction and $L_0 \equiv\sqrt{2A_{\text{ex}}/K_d}$ in the $x$-direction; the ratio $\ell\equiv\Delta/L_0$ measures the anisotropy between these two scales. Time is measured in units of $t_0 = (1+\alpha^2)/(\gamma M_s N_n)$, and the applied field $H_{z}$ is expressed in units of the Walker field $H_{W}=\alpha M_s N_n/2$ in the DMI-free case~\cite{skaugen2022}, with $h\equiv H_{z}/H_{W}$. We further introduce the dimensionless parameter $\delta \equiv \pi D/(4K_d\Delta)$, which quantifies the DMI field $H_{D}=\pi D/2\Delta\mu_{0}M_{s}$ relative to the magnetostatic field $H_{K}=2K_{d}/\mu_{0}M_{s}$~\cite{thiaville2012}. With $\mathbf{q} \equiv (u,\phi)^T$, the nondimensionalized system reads
\begin{equation}
    \frac{\partial\mathbf{q}}{\partial t} = \frac{1}{2}
    \left[
        \mathbf{f}(\phi)
        + \mathcal{G}(\phi)\,\frac{\partial\mathbf{q}}{\partial x}
        + \mathcal{D} \,\frac{\partial^2\mathbf{q}}{\partial x^2}
    \right],
    \label{eq:dot_X}
\end{equation}
with
\begin{align}
    \mathbf{f}(\phi) &\equiv
    \begin{pmatrix}
        \alpha^2 h + \sin 2\phi + 2\delta\cos\phi \\
        \alpha h - \alpha\sin 2\phi - 2\alpha\delta\cos\phi
    \end{pmatrix}, \label{eq:F}\\
    \mathcal{G}(\phi) &\equiv 2\ell \,g(\phi;\delta)
    \begin{pmatrix} 1 & \alpha \\ -\alpha & 1 \end{pmatrix}, \label{eq:A}\\
    \mathcal{D} &\equiv \begin{pmatrix} \alpha & -1 \\ 1 & \alpha \end{pmatrix}, \label{eq:B}
\end{align}
where $g$ is the scalar function
\begin{equation}
    g(\phi;\delta) \equiv \delta\sin\phi - \cos 2\phi.
    \label{eq:g}
\end{equation}
In the absence of DMI ($\delta=0$) and considering $\vert u_{x}\vert, \vert \phi_{x}\vert\ll1$, the model [Eqs.~\eqref{eq:dot_X}-\eqref{eq:B}] reduces to the one studied in Ref. \onlinecite{Pusiol2026}.

\subsection{The Walker model}

\begin{figure*}[t]
   \centering
   \includegraphics[]{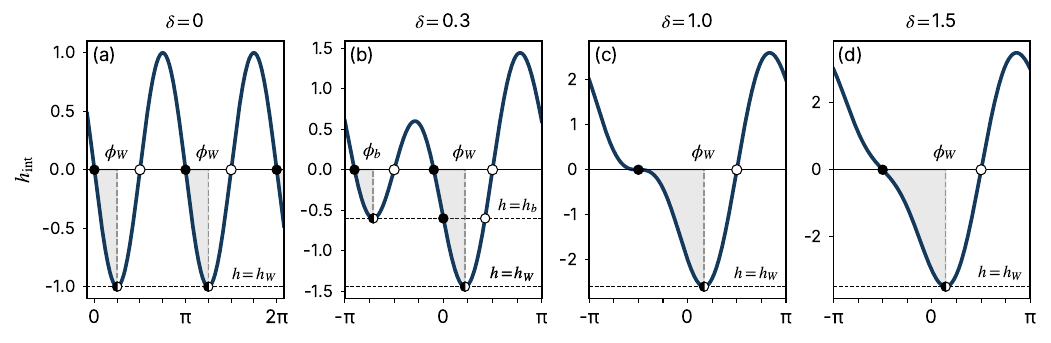}
   \caption{Internal field $h_{\text{int}}(\phi;\delta)$ for (a) $\delta=0$, (b) $\delta=0.3$, (c) $\delta=1$, and (d) $\delta=1.5$. Fixed points at applied field $h$ are the intersections of $h_{\text{int}}$ with the line $h_{\text{int}}=-h$. Black (white) dots mark stable (unstable) fixed points; at $h=0$ these are the roots of $h_{\text{int}}$. In every panel, the dashed line $h=h_W$ is tangent to the global minimum of $h_{\text{int}}$, located at $\phi_W$ (half-filled dot). Shaded areas indicate the stable branches. In panel (b), we additionally mark the secondary minimum at $\phi_b$ (half-filled dot), which defines the bifurcation curve $h_b(\delta)$. Two stable fixed points coexist for $h<h_b$. As $\delta\to1$ (c), $\phi_b$ merges with the stable root at an inflection point ($h_b\to0$), and for $\delta>1$ (d) only $\phi_W$ remains.}
   \label{Fig:f_phi_delta}
\end{figure*}

This paradigmatic model describes flat chiral DWs with a constant width $\Delta$, obtained from the more general description in Eqs.~\eqref{eq:phi} and \eqref{eq:theta} by restricting the collective coordinates to be spatially uniform~\cite{thiaville2012}, $u(x,t)\equiv u(t)$ and $\phi(x,t)\equiv\phi(t)$. The Walker model is conceptually important because it provides a simple framework for understanding the so-called Walker breakdown observed in micromagnetic simulations and experiments, and allows one to determine the characteristic Walker field separating the \textit{non-precessional} and \textit{precessional} regimes of DW motion. Since the original Walker model does not include DMI, we derive here its generalized form in the presence of DMI, which will also provide the basis for studying the spatial stability of the moving DW. The dynamical equations are obtained by setting all spatial derivatives to zero in Eq.~\eqref{eq:dot_X}, reducing the system to two coupled ordinary differential equations,
\begin{align}
    \dot{u} &= \frac{1}{2}\!\left[\alpha^2 h -h_{\text{int}}(\phi;\delta)\right],
    \label{eq:u_dot_1D}\\
    \dot{\phi} &= \frac{\alpha}{2}\!\left[h + h_{\text{int}}(\phi;\delta)\right],
    \label{eq:dot_phi_f}
\end{align}
where $h_{\text{int}}(\phi;\delta) \equiv -\sin 2\phi - 2\delta\cos\phi$ represents the internal DW field as the competition between the magnetostatic and the DMI field~\cite{Moretti2017}. Crucially, the $\phi$ equation is autonomous and decoupled from $u$: the internal angle $\phi(t)$ evolves independently, and the DW position $u(t)$ is then slaved to it through Eq.~\eqref{eq:u_dot_1D}. The non-precessional regime extends as long as Eq.~\eqref{eq:dot_phi_f} admits a fixed point $\phi^*$ with
\begin{equation}
    h + h_{\text{int}}(\phi^*;\delta) = 0.
    \label{eq:fix_f}
\end{equation}
This is possible only while $h$ does not exceed $|{\min_\phi h_{\text{int}}(\phi;\delta)}|$; the maximum such field defines the Walker field $h_W(\delta)$. The critical points of $h_{\text{int}}$ are
\begin{equation}
    \sin\phi_{\pm} = \frac{-\delta \pm \sqrt{\delta^2+8}}{4},
    \label{eq:critical_points}
\end{equation}
where the solution $\phi_{-}$ is valid for $\delta\leq1$, whereas $\phi_{+}$ is valid for $\delta\geq0$. One can show that $h_{\text{int}}(\phi_+;\delta)$ is the global minimum of $h_{\text{int}}$ for all values of $\delta$. The Walker field and corresponding magnetization angle are therefore
\begin{align}
    h_W(\delta) &= \frac{\left(3\delta+\sqrt{\delta^2+8}\right)
    \sqrt{4-\delta^2+\delta\sqrt{\delta^2+8}}}{4\sqrt{2}}, \label{eq:hW}\\
    \phi_W(\delta) &= \arcsin\!\left(\frac{-\delta+\sqrt{\delta^2+8}}{4}\right).
    \label{eq:phiW}
\end{align}

\subsubsection{Fixed Points, Stability, and Precession}
\label{sec:fixed_points}

In the non-precessional regime, the stable equilibria of Eq.~\eqref{eq:dot_phi_f} are the roots $\phi^*$ of the fixed-point condition Eq.~\eqref{eq:fix_f} that additionally satisfy
\begin{equation}
    g(\phi^*;\delta) < 0,
    \label{eq:fix_df}
\end{equation}
since $dh_{\text{int}}/d\phi|_{\phi^*} = 2g(\phi^*;\delta)$. The resulting fixed-point structure is shown in Fig.~\ref{Fig:f_phi_delta} and summarized in Table~\ref{tab:fixed:pts}: black (white) dots mark stable (unstable) solutions at $h=0$, and the shaded regions trace the stable branches as $h$ increases toward $\phi_W(\delta)$ or, where applicable, the secondary minimum $\phi_b(\delta)$.
\begin{table}[b]
\caption{\label{tab:fixed:pts} Stable fixed points in the non-precessional regime as a function of $\delta$ and the applied magnetic field $h$.}
\begin{ruledtabular}
\setlength{\extrarowheight}{2pt}
\begin{tabular}{lll}
 DMI strength & $h=0$ & $0<h< h_{W}$ \\ \colrule
 $\delta = 0$ & $\phi^{*}_{1} = 0$ & $0<\phi^{*}< \pi/4$ \\
  & $\phi^{*}_{2} = \pi$ & $\pi<\phi^{*}< 5\pi/4$ \\[2pt]
 \colrule
 $0<\delta<1$ & $\phi^{*}_{1} = \arcsin(\delta)-\pi$ & $\phi^{*}_{1}<\phi^{*}< \phi_{b}(\delta)$\footnote{$\phi_{b}(\delta)=\phi_{W}(-\delta) - \pi$, valid for $0<h\leq h_{b}(\delta)$.} \\[2pt]
  & $\phi^{*}_{2}=-\arcsin(\delta)$ & $\phi^{*}_{2}<\phi^{*}< \phi_{W}(\delta)$  \\[2pt]
 \colrule
 $\delta\geq1$ & $\phi^{*}_{1} = -\pi/2$ & $-\pi/2<\phi^{*}< \phi_{W}(\delta)$ \\[2pt]
\end{tabular}
\end{ruledtabular}
\end{table}
At $h=0$, for $\delta<1$ two stable fixed points flank the unstable point $\phi^*=-\pi/2$ [Figs.~\ref{Fig:f_phi_delta}(a)--(b)]; as $\delta\to1$ they merge continuously into the single Néel configuration that persists for $\delta\geq1$ [Figs.~\ref{Fig:f_phi_delta}(c)--(d)]. This is a supercritical pitchfork bifurcation~\cite{Strogatz2015} in $\delta$, occurring at the isolated point $(\delta,h)=(1,0)$.

For $0<h\leq h_W$, the field acts as an imperfection parameter, tilting this picture. Rather than track the fixed points individually, it suffices to note that each stable branch terminates where it collides with the neighboring unstable one, at the critical points of $h_{\text{int}}$ given by Eq.~\eqref{eq:critical_points}: $\phi_W(\delta)$ [global minimum, Eq.~\eqref{eq:phiW}] and, for $\delta<1$, the secondary minimum $\phi_b(\delta) = \phi_W(-\delta)-\pi$ [Fig.~\ref{Fig:f_phi_delta}(b)]. These are fold (saddle-node) points with $g(\phi;\delta)=0$.

The corresponding field threshold for $\phi_{b}(\delta)$ is,
\begin{equation}
    h_{b}(\delta) = \sqrt{\frac{4-\delta^2-\delta\sqrt{\delta^2+8}}{2}}
    \left(\frac{\sqrt{\delta^2+8}-3\delta}{4}\right).
    \label{eq:h_PB}
\end{equation}

Above the Walker field, no fixed point of Eq.~\eqref{eq:dot_phi_f} exists and the internal angle winds continuously with average
velocity $\langle\dot{\phi}\rangle = m\pi/T(h; \delta)$, where $T(h;\delta)$ is the field-dependent precession period and $m$ is the winding number per period: $m=1$ for $\delta=0$ and $m=2$ for
$\delta\neq 0$. The period is given by
\begin{equation}
    T(h;\delta) = \frac{2}{\alpha}
    \int_{0}^{m\pi} \frac{d\phi}{h + h_{\mathrm{int}}(\phi;\delta)}.
    \label{eq:period_general}
\end{equation}
For $\delta=0$ the integral~\eqref{eq:period_general} evaluates analytically to
\begin{equation}
    T(h;0) = \frac{2\pi}{\alpha\sqrt{h^{2}-1}},
    \label{eq:period_delta0}
\end{equation}
however, for $\delta\neq0$ it does not admit a compact closed form
for general $\delta$ and can be evaluated numerically. Finally, in this model the average velocity $v \equiv \langle \dot{u} \rangle$ of the DW is given by
\begin{equation}
v  =
    \begin{cases}
    \dfrac{1+\alpha^{2}}{2}h, \quad &\text{if}\; h\leq h_{W}   , \\
    \dfrac{1+\alpha^{2}}{2}h-\dfrac{m\,\pi}{\alpha T(h;\delta)}, \quad &\text{if}\;h>h_{W}, \\
    \dfrac{\alpha^{2}}{2}h, \quad &\text{if}\;h\gg h_{W}.
    \label{eq:u_dot_avg}
    \end{cases}
\end{equation}

\section{Stability of extended walls}
\label{sec:linearstability}

The results of Sec.~\ref{sec:reducedmodel} provide the basis for studying spatial instabilities of the moving DW. To determine the conditions under which a spatially uniform DW becomes unstable to small perturbations along the wall, we linearize Eq.~\eqref{eq:dot_X} around the corresponding uniform solution. We write
\begin{equation}
    \mathbf{q}(x,t) = \mathbf{q}_1(t) + \boldsymbol{\eta}(x,t),
    \qquad
    \boldsymbol{\eta}(x,t) \equiv \begin{pmatrix} Q(x,t) \\ \Phi(x,t) \end{pmatrix},
    \label{eq:linearization}
\end{equation}
where $\boldsymbol{\eta}$ is a small perturbation and $\mathbf{q}_1(t) = (u_1(t),\phi_1(t))^T$
is the flat-wall solution of Eqs.~\eqref{eq:u_dot_1D} and \eqref{eq:dot_phi_f}, which satisfies
\begin{equation}
    \partial_t \mathbf{q}_1 = \dfrac{\mathbf{f}(\phi_1)}{2}.
    \label{eq:flat_wall_eom}
\end{equation}
Since $\mathbf{q}_1$ is spatially uniform, $\partial_x\mathbf{q} = \partial_x\boldsymbol{\eta}$
and $\partial_{xx}\mathbf{q} = \partial_{xx}\boldsymbol{\eta}$.
Substituting into Eq.~\eqref{eq:dot_X} and expanding $\mathbf{f}(\phi)$ and $\mathcal{G}(\phi)$
to first order in $\Phi$,
\begin{align}
    \mathbf{f}(\phi) &\approx \mathbf{f}(\phi_1)
    + \frac{\partial\mathbf{f}}{\partial\phi}(\phi_1)\,\Phi + \mathcal{O}(\Phi^2),
    \label{eq:F_expand}\\
    \mathcal{G}(\phi) &\approx \mathcal{G}(\phi_1)
    + \frac{\partial \mathcal{G}}{\partial\phi}(\phi_1)\,\Phi + \mathcal{O}(\Phi^2),
    \label{eq:A_expand}
\end{align}
and discarding second-order terms $\mathcal{O}(\Phi\,\partial_x\boldsymbol{\eta})$,
the equation for the perturbation becomes
\begin{equation}
    \partial_t\boldsymbol{\eta} = \frac{1}{2}\left[
        \frac{\partial\mathbf{f}}{\partial\phi}(\phi_1)\,\Phi
        + \mathcal{G}(\phi_1)\,\partial_x\boldsymbol{\eta}
        + \mathcal{D}\,\partial_{xx}\boldsymbol{\eta}
    \right].
    \label{eq:linearized}
\end{equation}

To decouple the spatial modes, we apply the Fourier transform
$\boldsymbol{\eta}(x,t) \to \boldsymbol{\eta}_\kappa(t)$, with $\partial_x \to i\kappa$
and $\partial_{xx} \to -\kappa^2$, where $\kappa$ is the dimensionless wave vector
in units of $1/L_0$. Substituting into Eq.~\eqref{eq:linearized}, all modes decouple
and each satisfies
\begin{equation}
    \partial_t \boldsymbol{\eta}_\kappa = \frac{1}{2}\,M_\kappa(t)\,\boldsymbol{\eta}_\kappa,
    \label{eq:mode_eq}
\end{equation}
where the entries of the time-dependent dynamical matrix $M_{\kappa}(t)$ are
\begin{subequations}
\label{eq:Mkappa_elements}
\begin{align}
    M_{11} &= 2i\ell\kappa g - \alpha\kappa^2, \\
    M_{12} &= -2g + 2i\alpha\ell\kappa g + \kappa^2, \\
    M_{21} &= -2i\alpha\ell\kappa g - \kappa^2, \\
    M_{22} &= 2\alpha g + 2i\ell\kappa g - \alpha\kappa^2,
\end{align}
\end{subequations}
and the entire time dependence enters through the scalar function $g\equiv g[\phi_{1}(t);\delta]$. Since $\phi_{1}$ is a solution of the Walker model [Eq.~\eqref{eq:dot_phi_f}], we analyze $M_\kappa$ separately in each of the two dynamical regimes, the non-precessional and the precessional.

\subsection{Non-precessional regime}\label{Sec:stationary}

In this regime, we study the stability of each Fourier mode $\kappa$ considering the solutions described in Sec.~\ref{sec:fixed_points}. The dynamical matrix is therefore time-independent, and the stability of each mode reduces to a classical eigenvalue problem. The eigenvalues $\lambda_\pm$ of $M_\kappa$ are
\begin{equation}
    \lambda_\pm = \frac{\tau \pm \sqrt{\tau^2 - 4p}}{2},
    \label{eq:charpoly}
\end{equation}
where $\tau = \mathrm{tr}(M_\kappa)$ and $p = \det(M_\kappa)$. Stability of the flat wall requires $\mathrm{Re}(\lambda_\pm) < 0$, which is equivalent to the two conditions
\begin{align}
    &\text{(i)} \quad \tau_R \equiv \mathrm{Re}(\tau) < 0, 
    \label{eq:cond_i}\\
    &\text{(ii)} \quad \tau_R^2 > 
    \frac{\left|\tau^2 - 4p\right| + \mathrm{Re}\!\left(\tau^2 - 4p\right)}{2}.
    \label{eq:cond_ii}
\end{align}
Direct computation from Eq.~\eqref{eq:Mkappa_elements} yields
\begin{equation}
    \tau_R = 2\alpha\!\left(g - \kappa^2\right),
    \label{eq:trace_real}
\end{equation}
and condition (ii) implies
\begin{equation}
    \kappa^2 > 2g + 4\ell^2g^2.
\label{eq:cond_ii_final}
\end{equation}
The solutions $\phi_{1}$ with $g(\phi_{1};\delta) =0$ obey both conditions; therefore, they remain stable in this regime. The stable fixed points satisfy Eq.~\eqref{eq:trace_real} for all modes since $g<0$, however Eq.~\eqref{eq:cond_ii_final} imposes the following two regimes:
\begin{equation}
    \begin{cases}
        \ell^2 \leq \dfrac{1}{2|g|}: & \text{all }\kappa\geq0\text{ are stable}, \\[8pt]
        \ell^2 > \dfrac{1}{2|g|}: & \text{stable only for }\kappa > \kappa_\Lambda(g,\ell) \equiv \sqrt{4\ell^2g^2-2|g|}.
    \end{cases}
    \label{eq:kappa_cases}
\end{equation}
where $\kappa_{_\Lambda}$ is the critical wave vector, and the critical ratio is
\begin{equation}
    \ell_{c}(g)= \dfrac{1}{\sqrt{2\vert g \vert}}.
    \label{eq:ell_c}
\end{equation}

Conditions given by Eq.~\eqref{eq:kappa_cases} establish that DWs with length $L<2\pi/\kappa_{\Lambda}$ are stable for all $\ell$; otherwise, the stability of a mode $\kappa$ depends on the material-dependent ratio 
\begin{equation}
    \ell = \dfrac{\Delta}{L_{0}}=\sqrt{\dfrac{K_{d}}{2K_{0}}}=\sqrt{\dfrac{N_{n}}{2\tilde{K}}},
\end{equation}
with $\tilde{K} = 2K_{0}/\mu_{0}M^{2}_{s}$ as the effective quality factor~\cite{Malozemoff1979} measuring the strength of the effective anisotropy field in units of the demagnetizing field. Moreover, for thin films, the demagnetizing factor to lowest order in the film thickness $dz$ is~\cite{skaugen2019}
\begin{equation}
    N_{n} = \dfrac{dz}{\Delta}\dfrac{\ln2}{\pi}.
    \label{eq_Nn}
\end{equation}
Combined measured $M_{s}$, $K_{0}$ and $A_{\text{ex}}$ from Refs.~\onlinecite{krizakova2019, Pham2016}
yields $\ell\sim 0.1-0.3$ for all the reported samples, confirming $\ell\ll1$ as the physically relevant regime. Therefore, we analyze the most restrictive case by computing the absolute minimum $\ell^{\text{min}}_{c}$ of Eq.~\eqref{eq:ell_c}. If a sample is such that $\ell<\ell^{\text{min}}_{c}$, the flat moving DW is stable through the non-precessional regime. Since $g$ is bounded in the interval of the stable fixed points, the minimum value of $\ell_{c}$ is obtained by minimizing $g$
\begin{equation}
    g^{\min}(\delta) =
    \begin{cases}
        -\dfrac{\delta^{2}}{8}-1, \quad 0<\delta\leq4 \\
        1 - \delta, \quad \delta>4.
    \end{cases}
    \label{eq:g_min}
\end{equation}

\begin{figure}[]
    \centering
    \includegraphics[]{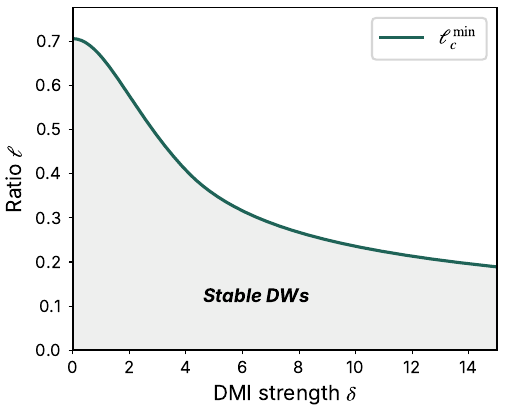}
    \caption{Absolute minimum of the critical ratio $\ell_{c}$ obtained analytically substituting Eq.~\eqref{eq:g_min} in Eq.~\eqref{eq:ell_c}.
    }
    \label{fig:lc_min}
\end{figure}

Figure~\ref{fig:lc_min} shows the absolute minimum of the critical ratio $\ell_{c}$ as a function of $\delta$. For moderate values of $\delta$, the ratio $\ell$ corresponding to the samples considered here lies below this absolute minimum, indicating that the flat DW is stable throughout the non-precessional regime. For sufficiently strong DMI, however, unstable modes can also develop in the non-precessional regime, leading to intricate DW structures at long times beyond the linear-response regime; an example is discussed in Appendix~\ref{app:nonprecessional_unstable}. In the following, we restrict our analysis to the parameter regime in which flat DWs remain stable throughout the non-precessional regime, and focus on the instabilities that emerge after Walker breakdown.

\subsection{Precessional regime}

\begin{figure*}[t]
   \centering
   \includegraphics[]{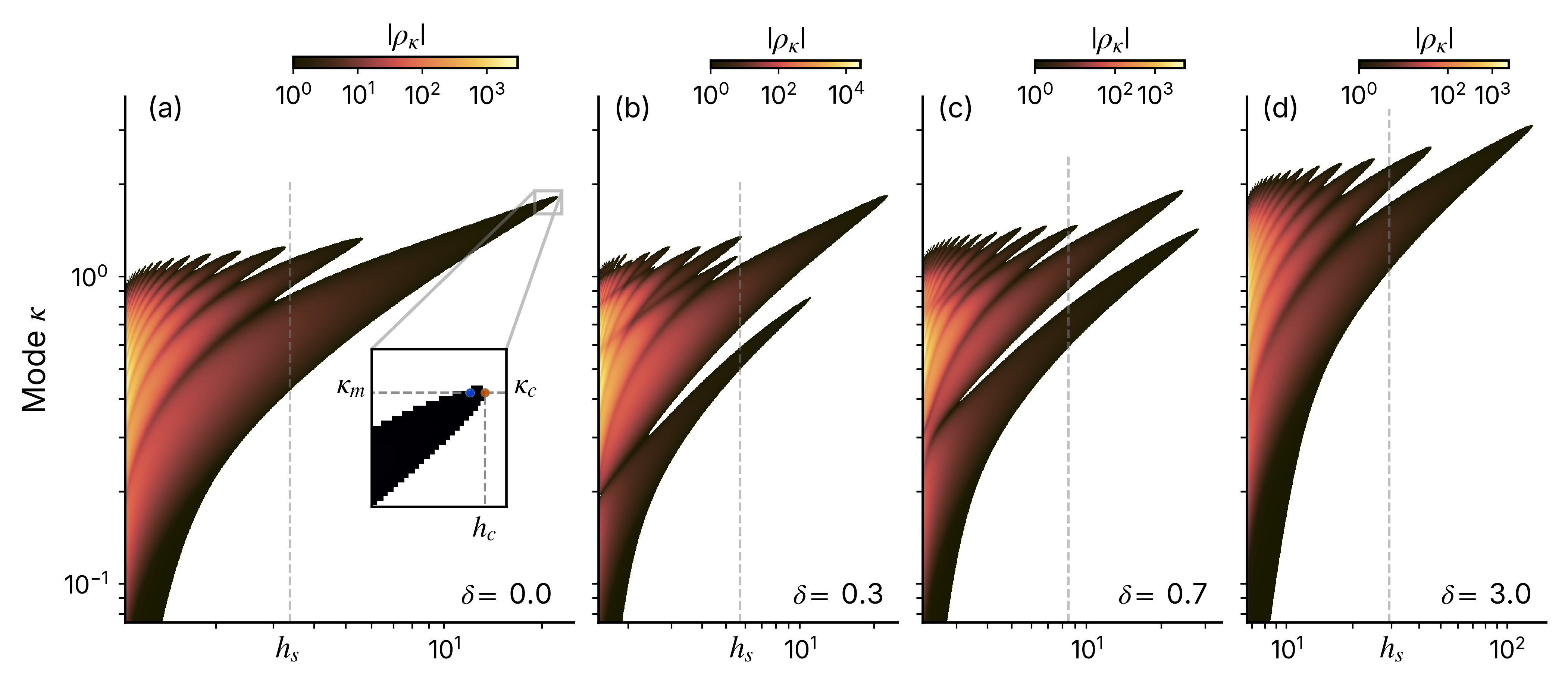}
   \caption{Stability diagrams in the $(h,\kappa)$ plane for $\alpha=0.15$ and $\ell=0.2$, at four values of the DMI strength: (a) $\delta=0$, (b) $\delta=0.3$, (c) $\delta=0.7$, (d) $\delta=3.0$. Color encodes the magnitude of the Floquet multiplier $|\rho_\kappa|$; white regions are linearly stable, while colored regions are unstable, with brighter shading indicating stronger instability. Vertical dashed lines mark the Slonczewski field $h_s$ [minimum value of Eq.~\eqref{eq:u_dot_avg}]. The inset in (a) shows a magnified view of the instability boundary near its termination, defining the critical field $h_c$, the maximum field at which an instability develops, and the associated critical mode $\kappa_c$; $\kappa_m$ denotes the maximum unstable wavevector, bounding the instability region from above.}

   \label{Fig:LS_alpha}
\end{figure*}

Above the Walker field, the flat-wall solution satisfies $\dot{\phi}_1 \neq 0$, so $\phi_1(t)$ evolves periodically in time and $M_\kappa(t)$ becomes a time-periodic dynamical matrix. The stability analysis of the stationary regime no longer applies, and a more general framework is required. For a linear system  with time-dependent periodic coefficients with period $T$ Floquet theory~\cite{Chicone2006} guarantees that the fundamental matrix solution $\Psi(t)$ admits the decomposition
\begin{equation}
    \Psi(t) = \Pi(t)\,e^{Ct},
    \label{eq:floquet_decomp}
\end{equation}
where $\Pi(t)$ is $T$-periodic and $C$ is a constant matrix. Asymptotic stability is then determined by the eigenvalues of the monodromy matrix $\mathcal{M} \equiv \Psi(T)$, known as the \textit{Floquet multipliers} $\rho_{i}$. Specifically, all solutions decay if and only if $|\rho_{i}| < 1$ for all eigenvalues.

In our system, since $\dot{\phi}_1 > 0$ for $h > h_W$, it is convenient to reparametrize time via $\phi_1$ \cite{Pusiol2026}, rewriting the linearized 
system~\eqref{eq:mode_eq} as
\begin{equation}
    \frac{d\boldsymbol{\eta}_\kappa}{d\phi_1} 
    = N_\kappa(\phi_1)\,\boldsymbol{\eta}_\kappa,
    \qquad
    N_\kappa(\phi_1) = \frac{M_\kappa(\phi_1)}{2\dot{\phi}_1}.
    \label{eq:reparam}
\end{equation}
Since $N_\kappa(\phi_1)$ is periodic in $\phi_1$, Floquet theory applies with the integration interval frozen to $\phi_1 \in [0, m\pi]$ ($m=1$ for $\delta=0$ and $m=2$ for $\delta\neq0$), regardless of the $h$-dependent physical period $T$. The monodromy matrix $\mathcal{M}_\kappa$ is obtained by integrating Eq.~\eqref{eq:reparam} over one period with initial conditions $\boldsymbol{\eta}_\kappa(0) = (1,0)$ and $(0,1)$, yielding its first and second columns respectively. A chiral flat DW is then linearly unstable for mode $\kappa$ if any Floquet multiplier satisfies $|\rho_\kappa| > 1$. The stability is analyzed by mapping $|\rho_\kappa|$ across the $(h, \kappa)$ plane for fixed $\alpha$ and $\delta$. 

Figure~\ref{Fig:LS_alpha} shows stability diagrams for different values of the parameter $\delta$, with fixed damping $\alpha = 0.15$ and $\ell = 0.2$. These parameters correspond to those of Pt/Co/GdOx~\cite{krizakova2019}, allowing direct comparison with previously reported numerical and experimental results for this system. Unstable regions ($|\rho_{\kappa}| > 1$) are colored according to the magnitude of the largest Floquet multiplier, while white regions denote stability
($|\rho_{\kappa}| < 1$). For any value of  $\delta$, the instability is bounded by a \textit{maximum wave vector} $\kappa_{m}$ and a \textit{critical field} $h_{c}$, as indicated in the inset of Fig.~\ref{Fig:LS_alpha}(a): flat DWs are linearly stable for all modes $\kappa$ whenever $h > h_{c}$, and unconditionally stable for systems of length $L < L_{m} \equiv 2\pi/\kappa_{m}$ for any $h > h_{W}$. For DWs longer than $L_{m}$, however, the allowed Fourier modes—the fundamental and its harmonics—each acquire a finite window of fields over which they are unstable. In the thermodynamic limit $L \to \infty$, the entire field range $h_{W}(\delta) < h \leq h_{c}$ is linearly unstable, as the continuous spectrum of wave vectors guarantees that at least one unstable mode exists at every field in this interval. Moreover, the instability field range extends beyond the negative differential mobility region bounded by $h_{S}$ computed from the minimum average velocity [Eq.~\eqref{eq:u_dot_avg}]. 

\subsubsection{Fast flow regime approximation}

The extent of the stability maps determines the characteristic length and field scales of the instability. In the following, we present an analytical approximation of $h_c$ as a function of both $\delta$ and $\alpha$. In the precessional regime, the angle $\phi_1(t)$ advances according to Eq.~\eqref{eq:dot_phi_f}. Because $\dot{\phi}_{1}$ is itself periodic in $\phi_1$, this advance is non-uniform, and this non-uniformity precisely generates the full comb of instability regions in the $(h, \kappa)$ plane. At large fields $h\gg h_{W}$, however, $\dot{\phi}_{1}\to \alpha h/2$, so that
\begin{equation}
\phi_1(t) \approx \phi_1(0) + \Omega t, \quad\Omega \equiv \frac{\alpha h}{2} = \text{const.}
\label{eq:phi_approx}
\end{equation}
Since the critical field $h_c$ is controlled by the outermost instability region, which occurs at large $h$ with respect to $h_{W}$ (See Fig.~\ref{Fig:LS_alpha}), this is precisely the regime in which the approximation is most reliable.

The angle $\phi_1$ enters the dynamical matrix $M_\kappa$ [Eq.~\eqref{eq:Mkappa_elements}] only through $g[\phi_1(t);\delta]$, so we rewrite the dynamical matrix for any $\phi_1(t)$ as
\begin{equation}
M_\kappa(\phi_1) = \langle M_\kappa \rangle + g[\phi_1(t);\delta]\,\Gamma_\kappa,
\end{equation}
with
\begin{align}
\langle M_\kappa \rangle &\equiv  \kappa^2\begin{pmatrix} -\alpha & 1 \\ -1 & -\alpha \end{pmatrix}, \\ 
\Gamma_\kappa &\equiv \begin{pmatrix} 2i\ell\kappa & -2(1-i\alpha\ell\kappa) \\ -2i\alpha\ell\kappa & 2(\alpha+i\ell\kappa) \end{pmatrix},
\end{align}
where $\langle {M}_\kappa \rangle$ is the constant background and $\Gamma_\kappa$ is a constant matrix. Under the high-field approximation,
\begin{equation}
g[\phi_1(t);\delta] \approx \delta\sin(\Omega t) - \cos(2\Omega t) , 
\end{equation}
the linearized dynamics therefore reduce to
\begin{equation}
\partial_{t}\boldsymbol\eta = \frac12 \big\{\langle {M}_\kappa \rangle + \Gamma_\kappa\big[\delta\sin(\Omega t) - \cos(2\Omega t)\big]\big\}\boldsymbol\eta,
\end{equation}
a constant background driven by a genuine two-harmonic periodic perturbation,
\begin{equation}
    \delta\sin(\Omega t) - \cos(2\Omega t)=\sum_{n}g_{n}e^{in\Omega t}
    \label{eq:Fourier_driving}
\end{equation}
with coefficients
\begin{equation}
    g_{\pm1}=\mp\dfrac{i\delta}{2}, \quad g_{\pm2}=-\dfrac{1}{2}.
    \label{eq:coeff}
\end{equation}
Since $\langle M_\kappa\rangle$ is constant, its contribution to the dynamics can be removed by moving to an interaction picture, $\boldsymbol\eta(t) \equiv e^{\langle M_\kappa\rangle t/2}\,\mathbf a(t)$, under which the equation of motion becomes
\begin{equation}
\partial_{t}\mathbf{a} = \frac12\,\overline{\Gamma}_\kappa(t)\,\sum_{n}g_{n}e^{in\Omega t}\,\mathbf a
\label{eq:a_eom}
\end{equation}
with $\overline{\Gamma}_\kappa(t) \equiv e^{-\langle M_\kappa\rangle t/2}\,\Gamma_\kappa\,e^{\langle M_\kappa\rangle t/2}$. Evaluating $\overline{\Gamma}_\kappa(t)$ explicitly,
\begin{equation}
\overline\Gamma_\kappa(t) = \Gamma_\kappa^{(0)} + \Gamma_\kappa^{(+)}e^{i\kappa^2t} + \Gamma_\kappa^{(-)}e^{-i\kappa^2t},
\label{eq:Gamma_rotated}
\end{equation}
with
\begin{align}
\Gamma_\kappa^{(0)} &= \begin{pmatrix} \alpha+2i\ell\kappa & -1+2i\alpha\ell\kappa \\ 1-2i\alpha\ell\kappa & \alpha+2i\ell\kappa \end{pmatrix}, \\[4pt]
\Gamma_\kappa^{(\pm)} &= \frac12\begin{pmatrix} \mp\alpha - i & -1 \pm i\alpha \\ -1 \pm i\alpha & \pm\alpha + i \end{pmatrix}.
\end{align}
The static term $\Gamma_\kappa^{(0)}$ carries no time dependence and cannot, by itself, resonate with the drive; only the rotating terms $\Gamma_\kappa^{(\pm)}e^{\pm i\kappa^2t}$, oscillating at the single frequency $\kappa^2$, can match a harmonic of Eq.~\eqref{eq:Fourier_driving} and produce secular growth.

Substituting Eqs.~\eqref{eq:Gamma_rotated} and \eqref{eq:Fourier_driving}--\eqref{eq:coeff} into Eq.~\eqref{eq:a_eom}, every term in the resulting sum oscillates at one of the frequencies $n\Omega$, $n\Omega+\kappa^2$, or $n\Omega-\kappa^2$. The static contribution ($n\Omega$) never vanishes, since $n=\pm1,\pm2$; the rotating contributions become secular precisely when
\begin{equation}
\kappa^2 = \Omega \quad (n=\pm1), \qquad\quad \kappa^2 = 2\Omega \quad (n=\pm2),
\label{eq:resonance}
\end{equation}
defining two distinct instability channels. Retaining only the secular term at each resonance and discarding the remaining, purely oscillatory contributions, the dynamics reduce to
\begin{equation}
\partial_t\mathbf a \approx \frac12\Big[g_n\,\Gamma_\kappa^{(-)} + g_{-n}\,\Gamma_\kappa^{(+)}\Big]\mathbf a, \quad n=1,2.
\label{eq:secular}
\end{equation}
The solution is a superposition of two normal modes $e^{\pm\nu_n t}$, with
\begin{equation}
    \nu_n = \frac{c_n}{4}\sqrt{1+\alpha^2},
    \label{eq:nu_n}
\end{equation}
where $c_1=\delta$, $c_2=1$. Reassembling the full solution via $\boldsymbol\eta(t)=e^{\langle M_\kappa\rangle t/2}\mathbf a(t)$, and using that both eigenvalues of $\langle M_\kappa\rangle$ share the same real part, $\operatorname{Re}\Lambda_\pm=-\alpha\kappa^2$, the dominant long-time behavior of $\boldsymbol\eta$ in each channel is set by a single effective rate,
\begin{equation}
|\boldsymbol\eta(t)| \sim \exp\!\left[\left(-\frac{\alpha\kappa^2}{2}+\operatorname{Re}\nu_n\right)t\right].
\label{eq:eta_rate}
\end{equation}
Instability therefore requires the resonant growth rate to exceed the background decay set by $\langle M_\kappa\rangle$, and the marginal (threshold) condition, defining the critical field $h=h_c^{(n)}$ for that channel, follows directly by equating the two:
\begin{equation}
\operatorname{Re}\nu_{n} = \frac{\alpha\kappa^2}{2}=\frac{n\alpha^{2}h^{(n)}_{c}}{4}.
\label{eq:threshold}
\end{equation}
The critical field is obtained by taking the larger of the two channels,
\begin{equation}
    h_{c}(\alpha, \delta)=\frac{\sqrt{1+\alpha^{2}}}{\alpha^{2}}\max\!\left(\delta,\frac12\right).
    \label{eq:hc_final}
\end{equation}
\begin{figure}[]
    \centering
    \includegraphics[]{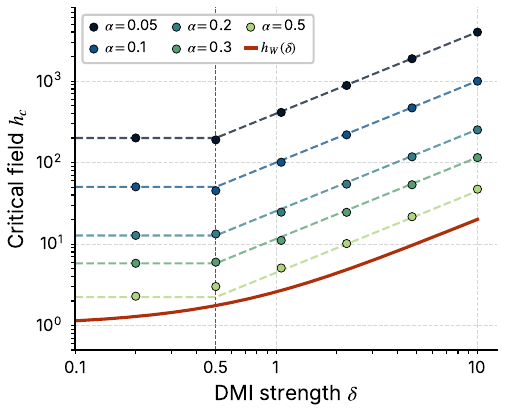}
    \caption{Critical field $h_c$ as a function of DMI strength $\delta$, for damping constants $\alpha=0.05,0.1,0.2,0.3,0.5$ (dark to light). Dots mark the maximum instability field extracted from exact Floquet-multiplier stability maps; dashed lines show the fast-flow approximation, Eq.~\eqref{eq:hc_final}. The solid red curve is the Walker field $h_W(\delta)$, shown for reference. The vertical dashed line at $\delta=0.5$ marks the crossover from the magnetostatic term to the DMI-driven term as the dominant contribution.
    }
    \label{fig:hc_delta}
\end{figure}

\begin{figure}[]
    \centering
    \includegraphics[]{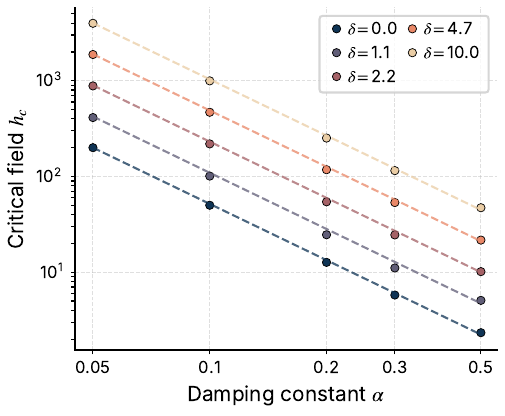}
    \caption{Critical field $h_c$ as a function of damping constant $\alpha$, for DMI strengths $\delta=0,1.1,2.2,4.7,10$. Dots mark the maximum instability field extracted from exact Floquet-multiplier stability maps; dashed lines show the fast-flow approximation, Eq.~\eqref{eq:hc_final}. All curves follow the $h_c\sim\sqrt{1+\alpha^{2}}/\alpha^{2}$ scaling predicted analytically, with the DMI strength setting the overall prefactor.
    }
    \label{fig:hc_alpha}
\end{figure}

A comparison between the critical field predicted by the fast-flow approximation, Eq.~\eqref{eq:hc_final}, and the values extracted from exact Floquet-multiplier stability maps is shown in Figs.~\ref{fig:hc_delta} and~\ref{fig:hc_alpha}. Dashed lines mark the analytical estimate; dots mark the maximum instability field obtained by mapping $|\rho_\kappa|$ over the $(h,\kappa)$ plane. At fixed $\alpha$, $h_c(\delta,\alpha)$ shows two regimes with DMI strength (Fig.~\ref{fig:hc_delta}): a plateau at $h_c(0,\alpha)$ for $\delta\lesssim1/2$, crossing over to linear growth in $\delta$ for $\delta\gtrsim1/2$. This sharp transition reflects the competition between the two resonant instability channels: the DMI-driven term $\propto\delta\sin\Omega t$ dominates for $\delta>1/2$, while the magnetostatic term $\propto\cos2\Omega t$ dominates for $\delta<1/2$, with the crossover located exactly at $\delta=1/2$ by the equal-height condition on the two channels. The analytic prediction agrees with the numerical threshold across the tested range of $\alpha$ and $\delta$, tightening as $\alpha\to0$. The only visible departure is at $\alpha=0.5$, where the crossover shifts to a somewhat lower $\delta$ than predicted. Fig.~\ref{fig:hc_alpha} confirms both the $\sqrt{1+\alpha^{2}}/\alpha^{2}$ scaling and the full quantitative prediction of Eq.~\eqref{eq:hc_final}.

\section{Phase diagram}
\label{sec:phasediagram}

\begin{figure}[t]
    \centering
    \includegraphics[]{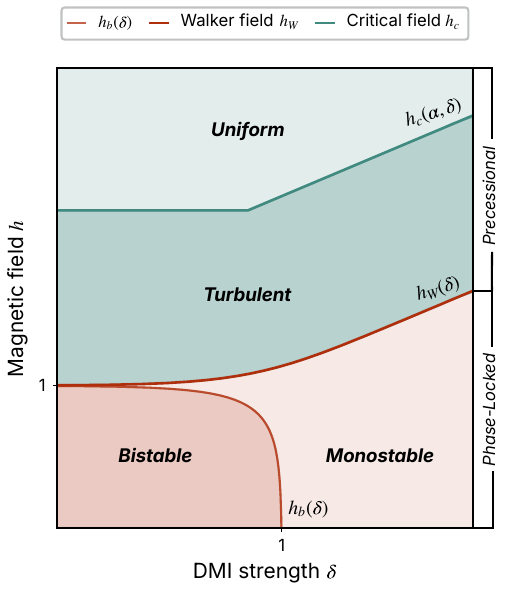}
    \caption{Phase diagram of the extended chiral DW dynamics model in the $(\delta, h)$ plane (logarithmic scales). The phase boundaries $h_W(\delta)$, $h_c(\alpha,\delta)$, and $h_b(\delta)$ separate the distinct dynamical regimes. The Walker field $h_W(\delta)$ marks the upper limit of the Phase-Locked regime, where $\dot{\phi}=0$; within this regime, $h_b(\delta)$ separates a Bistable region, in which the internal angle $\phi$ admits two solutions, from a Monostable region. Above $h_W(\delta)$, the system enters the Precessional regime ($\dot{\phi}\neq0$), which is further divided by the critical field $h_c(\alpha,\delta)$ into a Turbulent regime and a Uniform precession regime.}
    \label{fig:phases}
\end{figure}

We are now in a position to summarize all our results in a phase diagram indicating the dynamical regimes inferred from the spatial linear stability analysis of moving DWs and full non-linear micromagnetic simulations. Figure~\ref{fig:phases} shows the location of these different dynamical regimes in the parameter space spanned by the external magnetic field $h$ and the dimensionless DMI strength $\delta$.

We classify the dynamical regimes observed in the phase diagram based on the motion of the internal phase angle $\phi(x,t)$ and the spatial uniformity of the wall profile $u(x,t)$:
\begin{enumerate}
    \item \textbf{Phase-Locked Bistable:} $\dot{\phi} = 0$, with two stable internal tilt angles and a spatially uniform profile ($\partial_x u = 0$, $\partial_x \phi = 0$). This regime exists for $h\leq h_b(\delta)$, with $h_b(\delta)$ given by Eq.~\eqref{eq:h_PB}. The DW velocity is $v = \mu_{<} h$, with $\mu_{<}=(1+\alpha
    ^{2})/2$.
    \item \textbf{Phase-Locked Monostable:} $\dot{\phi} = 0$, with a unique stable internal tilt angle and a spatially uniform profile ($\partial_x u = 0$, $\partial_x \phi = 0$).  
    This regime exists for $h_b(\delta)<h\leq h_W(\delta)$ if $\delta<1$, and for $h\leq h_{W}(\delta)$ for all $\delta\geq 1$ with $h_W(\delta)$ given by Eq.~\eqref{eq:hW}. The DW velocity is $v = \mu_{<} h$.
    \item \textbf{Turbulent Precessional:} $\langle \dot{\phi} \rangle \neq0$, characterized by spatio-temporally irregular precession and dynamic wall roughening ($\partial_x u \neq 0$, $\partial_x \phi \neq 0$) occurring for $h_W(\delta) < h < h_c(\delta;\alpha)$, with $h_c(\delta;\alpha)$ approximately given by Eq.~\eqref{eq:hc_final}. 
    Rough DWs may crossover from single or multivalued profiles by increasing $\delta$. 
    The velocity $v$ is a complicated nonlinear function of the field.
    \item \textbf{Uniform Precessional:} $\langle \dot{\phi} \rangle \neq0$ , characterized by coherent precession where both $u$ and $\phi$ remain spatially flat ($\partial_x u = 0$, $\partial_x \phi = 0$) for $h > h_c(\delta;\alpha)$. In the fast flow regime $h \gg h_c(\delta;\alpha)$, 
    $v = \mu_{>} h$, with $\mu_> = \alpha^{2}/2 < \mu_<$ due to the precession acting as a source of dissipation.
\end{enumerate}

\begin{table*}[t]
\centering
\caption{Classification and main properties of the dynamical regimes of field-driven extended domain walls.}
\label{tab:regimes_summary}
\renewcommand{\arraystretch}{1.3}
\setlength{\tabcolsep}{8pt}
\begin{tabular}{l l c c}
\hline\hline
\textbf{Regime}
&
\textbf{Internal Phase ($\phi$)}
&
\textbf{Spatial Structure}
&
\textbf{Main Features}
\\
\hline
\textbf{Phase-Locked Bistable}
&
$\dot{\phi}=0$ (2 solutions)
&
$\partial_x u = \partial_x\phi = 0$
&
\parbox[t]{6.5cm}{Two stable tilt angles; steady translation.}
\\[2mm]
\textbf{Phase-Locked Monostable}
&
$\dot{\phi}=0$ (1 solution)
&
$\partial_x u = \partial_x\phi = 0$
&
\parbox[t]{6.5cm}{One stable tilt angle; steady translation.}
\\[2mm]
\textbf{Turbulent Precessional}
&
$\langle \dot{\phi} \rangle \neq0$ 
(chaotic)
&
$\partial_x u,\partial_x\phi \neq 0$
&
\parbox[t]{6.5cm}{Rough wall; spatiotemporal chaos and non-monotonous velocity-field characteristics.}
\\[2mm]
\textbf{Uniform Precessional}
&
$\langle \dot{\phi} \rangle \neq0$ 
 (periodic)
&
$\partial_x u = \partial_x\phi = 0$
&
\parbox[t]{6.5cm}{Coherent precession; homogeneous translation.}
\\
\hline\hline
\end{tabular}
\end{table*}

The phase diagram and properties summarized in Table~\ref{tab:regimes_summary} apply to sufficiently large domain walls, for which the spatial flexural modes form a dense continuum. These results hold across the range of micromagnetic parameters for which the uniform moving DW remains stable below the Walker threshold $h_W$, allowing both $h_W$ and $h_b$ to be accurately estimated directly from the DMI-aware Walker model. In contrast, determining the upper boundary $h_c$ requires analyzing the spatial stability of the flexural modes using the full extended $u$--$\phi$ domain wall model in Eq.~\eqref{eq:dot_X}.

Finally, we emphasize that the phase diagram reports the asymptotic steady-state regimes obtained from initially flat domain wall profiles in both position $u(x,0)$ and internal phase angle $\phi(x,0)$. Nevertheless, the Turbulent Precessional regime appears to be a unique attractor of the dynamics, with DWs initialized from arbitrary profiles of $u(x,0)$ and $\phi(x,0)$ converging to the same turbulent state (see Appendix~\ref{app:quench}). A systematic analysis of nonlinear hysteresis and multistability in the other dynamical regimes is beyond the scope of the present work.

\section{Conclusions}
\label{sec:conclusions}
We have constructed a phase diagram characterizing the stability of field-driven chiral domain walls as a function of drive field and DMI strength. The diagram separates a phase-locked regime from a
precessional regime, with the boundary set by the Walker field, which increases monotonically with DMI strength. This result extends the
DMI-free case of Ref.~\onlinecite{Pusiol2026}, where the
phase-locked-to-precessional transition was established at $\delta=0$, to the full $(h,\delta)$ parameter space.

In the precessional regime, we find that flexural fluctuations of the wall couple resonantly to the field-driven precession of the internal magnetization angle, and this coupling drives an instability. The upper field boundary of the unstable region is set jointly by the Gilbert damping and the DMI strength, extending beyond the
Slonczewski field. Low-dissipation systems support wide instability windows, scaling as $\alpha^{-2}$, while increasing DMI strength raises the instability ceiling further, producing a crossover in the
field scale that governs its extent.

Within this unstable field window, micromagnetic simulations reveal spatio-temporally chaotic wall dynamics. The interface roughens, and we associate this turbulent behavior with the nucleation and
annihilation of topological defects within the wall. Depending on drive conditions, the corrugated interface is found to be either single- or multivalued. Above the upper instability boundary, the wall re-stabilizes into a flat interface undergoing coherent precession. The agreement between the micromagnetic simulations and the reduced-model stability analysis demonstrates that the onset of these complex dynamical states can be traced to the instability of the spatially uniform precessional solution.

An important direction for future work is to assess the robustness of this phase diagram under more realistic conditions. In particular, disorder and thermal fluctuations are expected to modify the stability boundaries and the resulting domain wall dynamics; their inclusion will be essential for establishing a quantitative connection with experimental results. It will also be interesting to examine the response to dynamic field sweeps, where hysteresis effects may arise.

\begin{acknowledgments}
E. F. P. acknowledges support from the Simons-Balseiro interchange program. E. F. P. and A. B. K. acknowledge hospitality at LPTMS-Orsay. A. B. K. further acknowledges hospitality at DQMP-Geneva as well as support from the CNRS IRP Project “Statistical Physics of Materials”. We thank Stewart Barnes, Thierry Giamarchi, and Vincent Jeudy for fruitful discussions. Numerical simulations were performed on the HPC cluster of the Gerencia de Física at CAB-CNEA (SNCAD, Argentina).\end{acknowledgments}

\appendix

\section{Chiral Instabilities in the non-precessional regime}
\label{app:nonprecessional_unstable}

In Sec.~\ref{Sec:stationary}, we established that the spatial linear stability of moving domain walls in the non-precessional regime holds provided that the conditions in Eq.~(\ref{eq:kappa_cases}) are satisfied. While these conditions are broadly fulfilled across standard material parameters, they can break down under extreme regimes, such as for exceptionally large DMI strengths $\delta$. 

\begin{figure}[h!]
    \centering
    \includegraphics[]{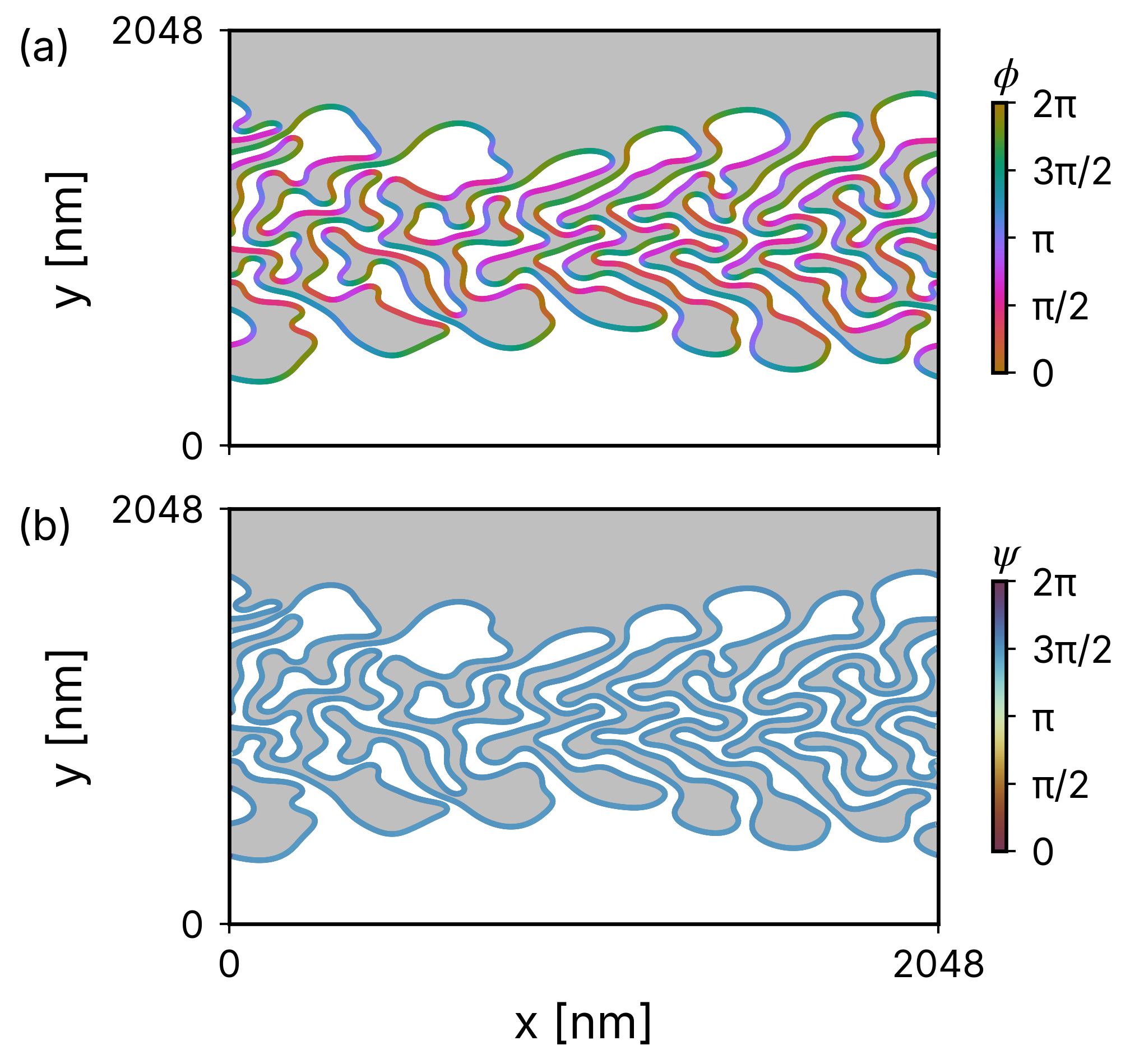}
    \caption{Magnetic configurations extracted from LLG simulations at $t=35$ ns in the non-precessional regime with $h=0$, starting from a slightly perturbed flat DW. (a) Interface colored by the internal magnetization angle $\phi$. (b) The same interface colored by the angle $\psi$, the angle of $\mathbf{m}$ relative to the local interface tangent. The uniform color in (b) shows a globally chiral structure.}
    \label{fig:labyrinth}
\end{figure}

Beyond the linear stability threshold, this instability drives the extended DW into intricate, labyrinthine patterns at long times, as illustrated in Fig.~\ref{fig:labyrinth} from full LLG micromagnetic simulations at $h=0$. We use the same material parameters as in Sec.~\ref{sec:mumaxphenomenology} but considering $D$ such as $\delta=12$.

Interestingly, in this regime the domain wall becomes spatially non-uniform in its position $u(s,t)$ while preserving the spatial uniformity of its internal phase $\psi(s,t)\equiv \psi$ (note that measuring the internal angle with respect to the local DW tangent, as shown in Fig.~\ref{fig:uphi}, is crucial to uncover this). As a result, the wall develops a corrugated yet globally chiral structure [see Fig.~\ref{fig:labyrinth}(b)]—in sharp contrast to the turbulent precessional regime, where phase coherence is lost. 

The formation of these structures reflects the competition between DMI and magnetostatic energy: when DMI dominates over exchange and anisotropy, the system lowers its total energy by swelling the domain wall perimeter through bending (flexural modes) while maintaining a fixed chiral internal phase alignment $\psi$.
\begin{figure}[]
    \centering
    \includegraphics[]{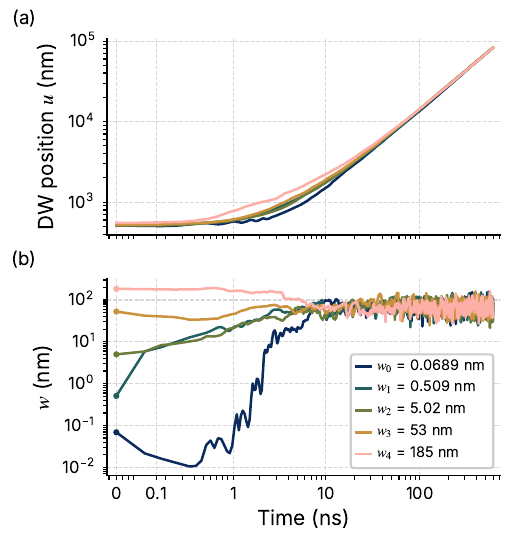}
    \caption{Dynamics of extended DWs starting from different initial conditions. (a) DW position $u(t)$ and (b) interface width $w(t)$ for five realizations with increasing initial width $w_i(0)$ (legend); all converge to the same long-time velocity and width statistics. The Turbulent Precessional regime is reached independently of the initial condition.}
    \label{fig:ini_test}
\end{figure}

\begin{figure}[b!]
    \centering
    \includegraphics[]{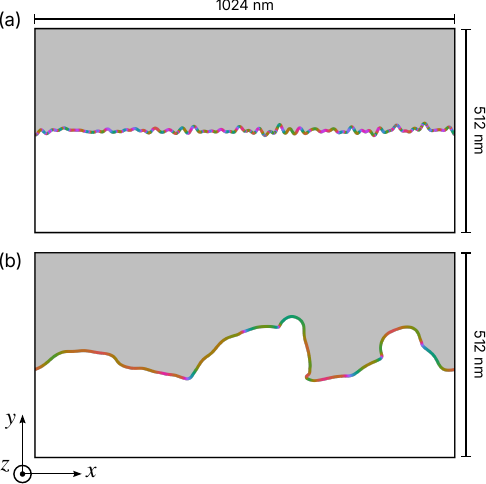}
    \caption{Magnetic configurations for initial condition $w_2$ of Fig.~\ref{fig:ini_test}. (a) Initial profile, $w_2(0) = 5.02\,\mathrm{nm}$. (b) Long-time configuration in the Turbulent Precessional regime.
    }
    \label{fig:ini_config}
\end{figure}

\section{Quenches to the Turbulent Precessional phase}
\label{app:quench}

The phase diagram constructed in Sec.~\ref{sec:phasediagram} describes the dynamical regimes of extended DWs starting from an initially flat interface with a homogeneous internal phase. We now show that the Turbulent Precessional regime is reached independently of the initial condition on $u(x,0)$ and $\phi(x,0)$. Micromagnetic simulations were performed with the same parameters as in Sec.~\ref{sec:mumaxphenomenology}, at the driving field $B_z = 71.4\,\mathrm{mT}$ [point (2) in Fig.~\ref{Fig:f_sec1}(a)]. The initial profiles were constructed as
\begin{align}
    u(x,0) &= A_{i}\sum_{n=504}^{511}\sin\left( \frac{2\pi n}{L_{x}}x+\beta_{n} \right),  \\
    \phi(x,0) &= B_{i}\sum_{n=0}^{50}\sin\left( \frac{2\pi n}{L_{x}}x+\gamma_{n} \right),
\end{align}
where $A_{i}, B_{i}$ are amplitudes that are constant within each realization, $L_{x}=1024\,\mathrm{nm}$, and $\beta_{n}, \gamma_{n}$ are independent random phases uniformly distributed in $[0,2\pi)$. 
Figure~\ref{fig:ini_test} shows, as a function of time, (a) the DW position $u$ and (b) the interface width $w$ [Eq.~\eqref{eq:width}] for five initial conditions spanning more than three decades in $w_{i}(0)$: from $w_0(0) = 0.0689\,\mathrm{nm}$, the near-flat case of Fig.~\ref{fig:transient}, to $w_4(0) = 185\,\mathrm{nm}$. As an example, the initial configuration of case $w_2$ is shown in Fig.~\ref{fig:ini_config}(a).

Despite this broad range of initial roughness, for all initial conditions tested, the dynamics converge to statistically equivalent long-time states, providing strong evidence for a robust turbulent attractor at this point of the phase diagram: the DW velocity approaches the same value regardless of initial condition [Fig.~\ref{fig:ini_test}(a)], and $w(t)$ fluctuates about a common mean and amplitude across all five realizations [Fig.~\ref{fig:ini_test}(b)]. The corresponding long-time magnetic configuration, exemplified in Fig.~\ref{fig:ini_config}(b), is likewise statistically indistinguishable across realizations despite the diversity of starting profiles.

\bibliography{refs}

\end{document}